\documentstyle[12pt]        {article}
\begin{document}
\title{Black Holes in 2+1 Teleparallel Theories
of Gravity}
\author{A.A. Sousa* and J.W. Maluf$^\dagger$ \\
\  Instituto de F\'{i}sica, Universidade de Bras\'{i}lia \\
C. P. 04385  \\
70.919-970 Bras\'{i}lia  DF   \\
Brazil}  
\maketitle

\begin{abstract}
We apply the Hamiltonian formulation of teleparallel theories of
gravity in 2+1 dimensions to a circularly symmetric geometry.
We find a family of one-parameter black hole solutions. The  BTZ
solution fixes the unique free parameter of the theory. The
resulting field equations coincide
with the teleparallel equivalent of Einstein's three-dimensional 
equations. We calculate the gravitational energy of the black holes
by means of the simple expression that arises in the Hamiltonian
formulation and conclude that the resulting value is identical to
that calculated by means of the Brown-York method.

\vfill

\noindent (*) e-mail: adellane@iesplan.br \par
\noindent ($\dagger$) e-mail: wadih@fis.unb.br\par
\end{abstract}
\newpage

\noindent{\bf \S 1. Introduction}\par
\bigskip
\bigskip

Gravity theories in three dimensions have gained considerable
attention in the last years \cite{1,2}. The expectation is that
the study of lower dimensional theories would provide relevant
information about the corresponding theory in four dimensions.
In 2+1 dimensions the Riemann and Ricci tensor have the same
number of components, consequently the imposition of Einstein's
equations in vacuum implies that the curvature tensor 
also vanishes. Therefore the space-time in the absence of
sources is flat and the existence of black hole would be
prevented\cite{3}. Moreover we know that Einstein's gravitational
theory in 2+1 dimensions does not have the Newtonian
limit\cite{3,4}. Therefore it would be desirable to establish a
theoretical formulation that would display the two important
features: the Newtonian limit and a black hole solution. 

In general relativity it is generally believed that the
gravitational energy cannot be localized. However, in the framework
of teleparallel theories gravity, as for instance the teleparallel
equivalent of general relativity (TEGR)\cite{5}, it is possible make
precise statements about the energy and momentum densities of
gravitational field. In the 3+1 formulation of the TEGR
it is found that the Hamiltonian and vector constraints contain
divergencies in the form of scalar and vector densities,
respectively. They can be identified as the energy and momentum
densities of the gravitational field\cite{6}. Therefore the
Hamiltonian and vector constraints are considered as energy-momentum
equations. This identification has proved to be consistent, and has
demonstrated that teleparallel theories
provide a natural instrument to the investigation of gravitational
energy. Several applications were presented in the literature,
among which we mention the analysis of gravitational energy of
rotating black holes\cite{7} and of Bondi's radiative
metric\cite{8}.

In the definition of the gravitational energy in 3+1 dimensions
a condition on the
triads is necessary: if we require the physical parameters of the
metric tensor such as mass and angular momentum to vanish, we
must have a vanishing torsion tensor, $T_{\left( i\right) jk}=0$.
Triads that yield a vanishing torsion tensor in any coordinated
system, after requiring the physical parameters of the metric
tensor to vanish, are called  reference space triads\cite{7}.

A family of three-parameter teleparallel theories  in 2+1
dimensions was proposed by Kawai and a Lagrangian formulation was
developed\cite{9}. With particular conditions on the parameters of
the theory, black hole solutions were found\cite{9}. These
solutions are quite different from the Schwarzschild and Kerr
black holes in $3+1$ dimensions. We also remark that black
hole-like solutions were discussed in Ref. \cite{10}.

In a previous paper we considered the canonical formulation of this
three-parameter theory\cite{11}, starting from the  Lagrangian
density of Ref. \cite{9}. Following Dirac's method\cite{12} and
requiring Schwinger's time gauge condition\cite{13} we found out
that the three-parameter family was reduced to just one-parameter
family. This unique family describes a theory with a first class
algebra of constraints. We concluded that the Legendre transform
is well defined only if certain conditions on the parameters were
satisfied. One of these conditions is $\frac{3}{4}%
c_{1}+c_{2}=0$\cite{11}. Kawai's conditions
($(3c_1+4c_2)\ne 0$ and $c_1(3c_1+4c_2)<0$) rule out the existence
of the Newtonian limit and of a general class of black holes in the
Hamiltonian formulation previously investigated. If the value of
the remaining free parameter is fixed as $c_{1}=-\frac{2}{3}$,
the final theory agrees with the three-dimensional Einstein's theory
in vacuum\cite{11}. However, it is possible to fix the value of
$c_{1}$ by means of an alternative physical requirement.

In this article we apply the canonical analysis developed in Ref. 
\cite{11} to a static, circularly symmetric geometry.
We introduce a cosmological constant and find
the well-known BTZ (Ban\~{a}dos, Teitelboim and Zanelli)
solution\cite{14}. This black hole solution has been used to study
quantum models in 2+1 dimensions, since  the corresponding
models in 3+1 dimensions are very complicated\cite{15}.

We also investigate the gravitational energy of the BTZ
solution in a region of radius $r_{0}$, and compare it with the
result obtained by means of the Brown-York method \cite{16}.
The results are identical. However, it will be noted that the result
obtained by means of our method is much simpler.

The article is divided as follows: in $\S 2$  we review the
Lagrangian and Hamiltonian formulation of the three-dimensional
gravitational teleparallelism. In $\S 3 $ we apply the
formalism to arbitrary gravitational  fields. In $\S 4$ we define 
the energy of gravitational field in 2+1 dimensions, and apply it
to the BTZ solution. Finally in $\S 5$ we present our
conclusions.

The notation  is the following: space-time
indices $\mu,$ $\nu $, ...
and global $SO(2,1)$ indices $a, b$,... run from  $0$ to $2$.
In the 2+1 decomposition latin indices indicate space indices
according to $\mu =0,i$ and $a=(0),(i)$. The flat
space-time metric is fixed for $\eta _{(0)(0)}=-1$.
\bigskip
\bigskip

\bigskip
\bigskip

\noindent{\bf \S 2. Hamiltonian formulation of gravitational
teleparallelism in 2+1 dimensions}\par
\bigskip

We introduce the three basic postulates that the
Lagrangian density for the gravitational field in empty space,
in the teleparallel geometry, must satisfy. It must be
invariant (i) under coordinate tranformations, (ii) under global
($SO(2,1)$) Lorentz's transformations, and (iii)  under parity
transformations. In the present formulation, we will add a
cosmological constant $\pm \Lambda =\mp \frac{2}{l^{2}}$ to the
Lagrangian density, where $\Lambda =-\frac{2}{l^{2}}$, is a
negative cosmological constant. The most general Lagrangian density
quadratic in the torsion tensor is written as

$$
L_{0}=e(c_{1}t^{abc}t_{abc}+c_{2}v^{a}v_{a}+
c_{3}a_{abc}a^{abc})\;.\eqno(2.1)  
$$
where $c_{1}$, $c_{2}$ and $c_{3}$ are constants,
$e=\det (e^a\,_\mu)$ and

$$
t_{abc}=\frac{1}{2}\left( T_{abc}+
T_{bac}\right) +\frac{1}{4}\left( \eta_{ca}v_{b}+
\eta _{cb}v_{a}\right) -\frac{1}{2}\eta _{ab}v_{c}\;.\eqno(2.2)
$$

$$
v_{a}=T^{b}{}_{ba}=T_a\;.\eqno(2.3)
$$

$$
a_{abc}=\frac{1}{3}\left( T_{abc}+
T_{cab}+T_{bca}\right)\;. \eqno(2.4)
$$

$$
T_{abc}=e_{b}{}^{\mu }e_{c}{}^{\nu }T_{a\mu \nu }\;.\eqno(2.5)
$$

The definitions above correspond the irreducible components
of the torsion tensor\cite{9}. With the purpose of obtaining the
Hamiltonian formulation, we will need to rewrite the three terms
of $L_{0}$ in a form such that the torsion tensor is factorized.
Thus  we rewrite $L_{0}$ as 

$$
L_{0}=e\left( c_{1}X^{abc}+c_{2}Y^{abc}+c_{3}Z^{abc}\right)
T_{abc}\;,\eqno(2.6)
$$
where

$$
X^{abc}=\frac{1}{2}T^{abc}+\frac{1}{4}T^{bac}-
\frac{1}{4}T^{cab}+\frac{3}{8}%
\left( \eta ^{ca}v^{b}-\eta ^{ba}v^{c}\right)\;,\eqno(2.7a)
$$

$$
Y^{abc}=\frac{1}{2}\left( \eta ^{ab}v^{c}-
\eta ^{ac}v^{b}\right)\;,\eqno(2.7b)
$$

$$
Z^{abc}=\frac{1}{3}\left( T^{abc}+T^{bca}+T^{cab}\right)
\;.\eqno(2.7c)
$$
The definitions above satisfy

$$
X^{abc}=-X^{acb}\;,\;\;Y^{abc}=-Y^{acb}\;,\;\;
Z^{abc}=-Z^{acb}\;.$$

We define the quantity $\Sigma ^{abc}$ according to

$$
\Sigma ^{abc}=c_{1}X^{abc}+c_{2}Y^{abc}+c_{3}Z^{abc}\;,
$$
which allows us rewrite $L_{0}$ as

$$
L_{0}=e\Sigma ^{abc}T_{abc}\;.
$$

The Hamiltonian formulation is obtained by writing the Lagrangian
density in first order differential form. We will consider the 
space-time to be empty. In order to carry out the Legendre
transform, we make a 2+1 decomposition of the space-time triads
$^{3}e^{a}{}_{\mu }$\cite{11}. Such decomposition  
projects $^{3}e^{a}{}_{\mu }$ to the
two-dimensional spacelike hypersurface. The Legendre transform
is implemented provided we eliminate some velocity
components\cite{11}. We find that two conditions are necessary
to implement the Legendre transform:

$$c_{2}+\frac{3}{4}c_{1}=0\;,\eqno(2.8a)$$

$$c_{1}=-\frac{8c_{3}}{3}\;.\eqno(2.8b)$$
With the addition of a cosmological
constant the Lagrangian density can be written as\cite{11}

$$
L =P^{\left( j\right) i} \dot{e}_{\left( j\right)
i}+N^{k}C_{k}+NC-\partial _{i}\left[ N_{k}P^{ki}+
N(3c_{1}eT^{i})\right] 
+\lambda ^{ij}P_{\left[ ij\right] }\;,\eqno(2.9)
$$
where $P^{(j)i}=\delta L/\delta \dot e_{(j)i}\;$,
$P^{ki}=e_{(j)}\,^kP^{(j)i}$ and
$T^i=g^{ik}T_k=g^{ik}e^{(j)m}T_{(j)mk}$.
Therefore the Hamiltonian density is given by

$$
H(e_{\left( j\right) i},P^{\left( j\right) i})=
-N^{k}C_{k}-NC+\partial _{i}
\left[ N_{k}P^{ki}+N(3c_{1}eT^{i})\right]
-\lambda ^{ij}P_{\left[ ij\right]}\;,\eqno(2.10)  
$$
where

$$
C_{k}=e_{\left( j\right) k}\partial _{i}P^{\left( j\right) i}
+P^{\left(
j\right) i}T_{\left( j\right) ik}\;,\eqno(2.11)
$$

$$
C=\frac{1}{6ec_{1}}\left( P^{ij}P_{ji}-P^{2}\right)
+eT^{ikj}\Sigma
_{ikj}-\partial _{k}\left[ 3c_{1}eT^{k}\right]
\pm e\frac{2}{l^{2}}\;,\eqno(2.12)
$$
are the vector and Hamiltonian constraints, respectively. $N^{k}$,
$N$ and $\lambda ^{ij}=-\lambda_{ji}$ are Lagrange multipliers,
and $\pm \frac{2}{l^{2}}=\mp
\Lambda $. The adition of $\Lambda $ to the Lagrangian
density affects only the Hamiltonian constraint $C$. The Legendre
transform reduced the three-parameter family to a one-parameter
family of theories.

A consistent implementation of the Legendre transform is a
necessary condition for the Hamiltonian formulation. However,
it is not sufficient. It is also necessary to verify whether
the constraints
are first class, i.e., if the algebra of constraints ``closes''.
Such analysis has been investigated in Ref.\cite{11}. The conclusion
is that all constraints of the theory are indeed first class.\par
\bigskip
\bigskip

\noindent {\bf \S 3. Applications of the Hamiltonian formulation
in 2+1 dimensions}\par
\bigskip
\noindent {\bf \S 3.1 The Newtonian limit}\par
\bigskip

The field equations for the Lagrangian density (2.1)
were obtained in Ref. \cite{9}.
It was noticed that these field equations are equivalent to the
three-dimensional Einstein's equations if the parameters are
fixed according to

$$ c_{1}+\frac{2}{3} =0\;,\eqno(3.1)$$

$$c_{2}-\frac{1}{2} =0\;,\eqno(3.2)$$

$$c_{3}-\frac{1}{4} =0\;.\eqno(3.3)$$

We know that the three-dimensional Einstein's equations do
not have the Newtonian limit and neither
black holes solutions. In Ref. $\cite{9}$ it was attempted to
obtain the Newtonian limit of the field equations,
considering static circularly symmetric fields. For this purpose
it was found a relation between the parameters
parameters $c_{1}$ and $c_{2}$ (in our notation),

$$
3c_{1}+4c_{2} =-6c_{1}c_{2}\;,\eqno(3.4a)$$

$$c_{1}c_{2} \neq 0\;.\eqno(3.4b)$$
Condition (2.8a), $c_{2}+\frac{3}{4}c_{1}=0$ obtained in the
Hamiltonian formulation violate the condition above.
Therefore the Newtonian limit is excluded from the theory defined
by (2.9) and (2.10).

In the investigation by Kawai\cite{9} it was found a condition on
the parameters of the theory that yields a black hole
solution. This condition is $c_{1}(3c_{1}+4c_{2})<0$
and refer to a particular solution (equation (5.15) of Ref. \cite
{9}) of the theory. Conditions $c_{2}+\frac{3}{4}c_{1}=0$
and $c_{1}=-\frac{8c_{3}}{3}$  for the Legendre transform
prevent black hole solutions. However, it is not excluded the
possibility that different solutions of Eq. (5.15) of
Ref. \cite{9} yield black hole solutions that violate the
condition $c_{1}(3c_{1}+4c_{2})<0$.

The Hamiltonian formulation above yields equations that are
precisely equivalent to the three-dimensional Einstein's equations
in vacuum, as it can be verified by fixing $c_{1}=-\frac{2}{3}$ in
equations (2.10) and (2.12). However, it is
worth investigating whether another value of $c_{1}$
is acceptable for a gravitational theory, since there does not 
exist a compelling physical reason for fixing $c_{1}=-\frac{2}{3}$.
We will address this issue in the next section.
\bigskip

\noindent{\bf \S 3.2 Black holes in the
circularly symmetric  geometry}\par
\bigskip

In this section we investigate the fixing of the free parameter
in the framework of an exact vacuum solution.
We will carry out a symmetry reduction and consider the constraints
in a circularly symmetric geometry. We will consider the following 
space-time metric that describes a rotating source,

$$ds^2=-N^2 dt^2+ f^{-2}dr^2+r^2(
N^\phi dt + d\phi)^2\;,
$$
where $f$, $N$ and $N^\phi$ are {\it a priori} independent functions
of $r$ and $t$. The diads for the spatial sector
of the metric tensor above are  given by

$$(e_{\left( k\right) i})=\left( 
\begin{tabular}{ll}
$f(r,t)^{-1}$ & $0$ \\
$0$ & $r$%
\end{tabular}
\right)\;.\eqno(3.5)
$$
$\left( k\right)$ and $i$ are row and column indices, respectively.

The two-dimensional metric $g_{ij}$ reads

$$ (g_{ij})=\left( 
\begin{tabular}{ll}
$f(r,t)^{-2}$ & $0$ \\ 
$0$ & $r^{2}$
\end{tabular}
\right)\;,\eqno(3.6)
$$
and $e=\det \left( e_{\left( k\right) i}\right) =rf(r,t)^{-1}$.

The nonzero components of the torsion tensor on the spacelike
section are
$T_{\left( 2\right) 12} =- T_{\left( 2\right) 21}= 1$.
With the help of the inverse metric tensor $g^{ij}$ we can write
the inverse diads $e_{\left( k\right) }^{i}$,

$$
(e_{\left( k\right) }\,^{i})=\left( 
\begin{tabular}{ll}
$f(r,t)$ & $0$ \\ 
$0$ & $\frac{1}{r}$%
\end{tabular}
\right)\;.
$$
It follows that

$$
T^{\left( 2\right) 12} =
g^{11}g^{22}T^{\left( 2\right)}\,_{12}
=\frac{f(r,t)^{2}}{r^{2}}\;,$$

$$
T^{\left( 2\right) 21} =
g^{11}g^{22}T^{\left( 2\right)}\,_{12}
=-\frac{f(r,t)^{2}}{r^{2}}\;.$$

Let us calculate the term $e\Sigma ^{kij}T_{kij}=
e\Sigma_{(k)(i)(j)}T^{(k)(i)(j)}$ of the Hamiltonian
constraint. It is not difficult to obtain

$$
e\Sigma _{kij}T^{kij} =
e\Sigma _{\left( k\right) ij}T^{\left( k\right) ij}
=2e\Sigma _{\left( 2\right) 12}T^{\left( 2\right) 12}\;,
$$
where

$$
\Sigma _{abc}=c_{1}X_{abc}+c_{2}Y_{abc}+c_{3}Z_{abc}\;,
$$
and
$$
e^{b}\,_{i} e^{c}\,_{j}
\Sigma _{abc}=\Sigma _{aij}.
$$
The latter quantity can  be written in terms of the torsion
tensor $T_{aij}$,

$$
\Sigma _{aij} =\left( \frac{c_{1}}{2}+\frac{c_{3}}{3}\right)
T_{aij}+\left( \frac{c_{1}}{4}-
\frac{c_{3}}{3}\right) \left(
e^{(k)}\,_i e_{a}\,^{l}
T_{\left( k\right) lj}-
e^{(k)}\,_j e_{a}\,^{l}T_{\left(
k\right) li}\right)
$$

$$
+\left( \frac{3}{16}c_{1}-\frac{c_{2}}{2}\right) \left(
e_{aj}T_{i}-e_{ai}T_{j}\right) ,
$$
where

\begin{eqnarray*}
T_{i} &=&T^{b}\, _{bi}=
e_{b}\,^{m}T^{b}\,_{mi} 
=e_{\left( k\right) }\,^{m}
T^{(k)}\,_{mi} ,
\end{eqnarray*}

After a long calculation we obtain

$$
e\Sigma _{kij}T^{kij} =
e\Sigma _{\left( k\right) ij}T^{\left( k\right) ij}
=e\left( c_{2}+\frac{3}{4}c_{1}\right)=0.
\eqno(3.7)$$

From the momentum $P^{\left( k\right) j}$,
canonically conjugated to $e_{\left( k\right) j}$,
it is possible to construct the
tensor density  $P^{ij}=e_{\left(k\right) }\,^{i}
P^{\left( k\right) j}$ of weight +1.
We will determine the most general form of the
tensor $M_{ij}=e^{-1}P_{ij}$ restricted to a
circularly symmetric geometry. For this purpose we 
consider the Killing vector,

$$(\xi_i)=( 0\;,\; br^{2}),\eqno(3.8)$$
where $b$ is an arbitrary constant, and
require that the Lie derivative of $M_{ij}$ vanish,

$$
L_{\xi }\left( e^{-1}P_{ij}\right) =0\;.$$

We find that the most general matrix for $P_{ij}$ is given by:

$$
(P_{ij})=\left( 
\begin{tabular}{ll}
$rf(r,t)^{-1}A(r,t)$ & $P_{12}(r,t)$ \\ 
$P_{21}(r,t)$ & $rf(r,t)^{-1}B(r,t)$%
\end{tabular}
\right)\;.
$$
where $A(r,t)$ and $B(r,t)$ are arbitrary functions.

We can now obtain the momentum components
$P^{(k)j}=e^{(k)}\,_i P^{ij}$ in the time gauge. They read

\begin{eqnarray}
P^{\left( 1\right) 1} &=&rf(r,t)^{2}A(r,t),  \nonumber \\
P^{\left( 2\right) 1} &=&rP^{21}, \nonumber \\ 
P^{\left( 1\right) 2} &=&f(r,t)^{-1}P^{12}, \nonumber \\ 
P^{\left( 2\right) 2} &=&\frac{f(r,t)^{-1}}{r^{2}}B(r,t). 
\nonumber 
\end{eqnarray}

We proceed to obtain the Hamiltonian in terms of the quantities
above. The term $P^{ij}P_{ji}$ is written as

$$
P^{ij}P_{ji}=r^{2}f^{2}A^{2}+f^{-2}\frac{1}{r^{2}}%
B^{2}+2P^{21}P_{21}\;,
$$
where we have considered $P^{21}=P_{12}.$
The scalar density $P$ reads

$$P=frA+f^{-1}\frac{1}{r}B\;.$$
Therefore

$$
P^{ij}P_{ji}-P^{2}=2P^{21}P_{21},\eqno(3.9)
$$
where we have made $B(r,t)=0$ because there is no momentum
canonically conjugated to $g_{\theta \theta }$. We also obtain

$$
\partial _{k}\left( eT^{k}\right) =-\partial _{r}f(r,t).
\eqno(3.10)
$$

The vector constraints are easily calculated,

\begin{eqnarray}
C_{1} &=&C_{r}=-f^{-1}\partial _{r}\left( f^{2}\bar{P}
\right) ,  \nonumber \\
C_{2} &=&C_{\theta }=-r\partial _{r}\left( rP^{21}\right)
-rP^{21},  \nonumber 
\end{eqnarray}
where $\bar{P}=-f^{-1}P$,
and the Hamiltonian constraint is reduced to

$$
C(r,t) =-\frac{f(r,t)}{6rc_{1}}\left[ 2P^{21}(r,t)
P_{21}(r,t)\right] 
-3c_{1}\partial _{r}f(r,t)\mp \frac{2}{l^{2}}f(r,t)^{-1}r,  
\eqno(3.11)
$$
where the minus sign in the last term of the above expression
refers to a negative cosmological constant.
Finally we notice that
$\lambda_{ij}P^{\lbrack ij\rbrack}=0$, since the matrix
$P^{ij}$ is  symmetric in the circularly symmetric geometry.

We may now construct the action integral. It is given by

$$
I = \int dtdrd\theta\, L= 2\pi \int dtdr\left( \bar{P}
\dot{f}-N^{i}C_{i}-NC\right)\;.\eqno(3.12)
$$

At $t=0$ the equation $C_{\theta }(r,0)=0$,

$$
r\partial _{r}\left[ rP^{21}(r,0)\right] +rP^{21}(r,0)=0
\;,\eqno(3.13)
$$
has a simple solution,

$$ P^{21}(r,0)=\frac{k}{r^{2}}\;,\eqno(3.14)$$
where $k$ is a constant of integration. We also have

$$P_{21}(r,0)=g_{22}g_{11}P^{21}(r,0)=k\,f(r,0)^{-2},$$
and

$$P_{2}^{1}=g_{22}P^{21}(r,0)=k=P_{\theta }^{r},$$
where $P_{\theta }^{r}$ is the canonical variable conjugated to
the cyclic variable $g_{12}$=$g_{r\theta }$. It is a conserved
quantity, since

$$
\dot{P}^{21}(r,t)=-\frac{\delta H}{\delta g_{21}(r,t)}=0.
$$
Therefore at any time $t>0$ we have

$$
P^{21}(r,t)=\frac{k}{r^{2}}.\eqno(3.15)
$$

By replacing Eq. (3.15) in the Hamiltonian constraint we
arrive at

$$
C(r,t) =-\frac{f^{-1}(r,t)}{3c_{1}r^{3}}k^{2}-
3c_{1}\partial _{r}f(r,t)\mp \frac{2}{l^{2}}f(r,t)^{-1}r.
\eqno(3.16)
$$

Let us consider now the equation $C(r,0)=0$. We have

$$
-3c_{1}\partial _{r}f(r,0)-\frac{2}{l^{2}}f(r,0)^{-1}r
-\frac{f^{-1}(r,0)}{%
3c_{1}r^{3}}k^{2}=0,\;\;\;\;\Lambda<0 \eqno(3.17)
$$

$$
-3c_{1}\partial _{r}f(r,0)+\frac{2}{l^{2}}f(r,0)^{-1}r
-\frac{f^{-1}(r,0)}{%
3c_{1}r^{3}}k^{2}=0,\;\;\;\;\Lambda >0 \eqno(3.18)
$$
where $\Lambda =\pm \frac{2}{l^{2}}$. The solutions are

$$
f(r,0)^{2}=m-\frac{2}{3c_{1}}\frac{r^{2}}{l^{2}}
+\frac{1}{9c_{1}^{2}}\frac{%
k^{2}}{r^{2}},\;\;\;\;\Lambda <0
$$

$$
f(r,0)^{2}=m+\frac{2}{3c_{1}}\frac{r^{2}}{l^{2}}
+\frac{1}{9c_{1}^{2}}\frac{%
k^{2}}{r^{2}},\;\;\;\;\Lambda >0
$$
where $m$ is constant of integration. It is not difficult to
conclude that these solutions are independent of time. The time
evolution of $f(r,t)$ is given by

$$
\dot{f}(r,t)=\frac{\delta H}{\delta \bar{P}(r,t)}=0,
$$
where we have fixed the coordinate system such that $N^{1}(r,t)=0$.
Therefore at any later time $t$ we have

$$
f(r)^{2}=m-\frac{2}{3c_{1}}\frac{r^{2}}{l^{2}}
+\frac{1}{9c_{1}^{2}}\frac{%
k^{2}}{r^{2}},\;\;\;\;\Lambda <0 \eqno(3.19)
$$

$$
f(r)^{2}=m+\frac{2}{3c_{1}}\frac{r^{2}}{l^{2}}
+\frac{1}{9c_{1}^{2}}\frac{%
k^{2}}{r^{2}}.\;\;\;\;\Lambda >0   \eqno(3.20)$$
When $c_{1}=-\frac{2}{3}$ Eq. (3.19) becomes

$$
f(r)^{2}=m+\frac{r^{2}}{l^{2}}
+\frac{k^{2}}{4r^{2}}.\;\;\;\;
\Lambda <0  \eqno(3.21)
$$

By identifying $m=-M$ and $k=J$ this solution is known in the
literature as the rotational black hole solution
in the three-dimensional space-time of Ban\~{a}dos, Teitelboim
and Zanelli (BTZ)\cite{14}, with negative cosmological constant. In
the latter work, Ban\~{a}dos {\it et al.} used a time independent
metric. At this point we will likewise restrict the considerations
to a static field configuration. Since the spacelike sector of the
space-time triads given by Eq. (3.5) is time independent, the
restriction to a static geometry amounts to further requiring the
lapse and shift functions to be time independent (note that we have
already required $N^1=0$). If the space-time geometry is static,
then we must have

$$
\dot{\bar P}=-\frac{\delta H}{\delta f(r)}=0.\eqno(3.22)
$$ 

Consequently we have

$$
-3c_{1}\partial _{r}N(r)\mp \frac{2}{l^{2}}
f(r)^{-2}rN(r)-\frac{%
f(r)^{-2} }{3c_{1}r^{3}}N(r)k^{2}=0,
\eqno(3.23)
$$
for a negative cosmological constant. Making use of
Eq. (3.19) we arrive at

$$
-3c_{1}\partial _{r}N(r)
-\frac{1}{\left( m-\frac{2}{3c_{1}}\frac{r^{2}}{%
l^{2}}+\frac{1}{9c_{1}^{2}}
\frac{k^{2}}{r^{2}}\right) }\left( \frac{2}{%
l^{2}}r+\frac{k^{2}}{3c_{1}r^{3}}\right) N(r)=0,
$$
from what we obtain a solution for $N(r)$,

$$
N(r)=K\sqrt{\left( m-\frac{2}{3c_{1}}
\frac{r^{2}}{l^{2}}+\frac{1}{9c_{1}^{2}%
}\frac{k^{2}}{r^{2}}\right) }=f(r),\eqno(3.24)$$
where $K$ is an arbitrary constant.
The result above is identical to the corresponding result of
Ref. \cite{14} provided we make $K=1$,
$c_{1}^{{}}=-\frac{2}{3}$ and identify $k=J$. We will require
the constant $K$ to satisfy $K=1$, which amounts to a
redefinition of the time scale of the theory (this redefinition
also implies $K=1$ in Eqs. (3.25) and (3.26) below).

For a positive cosmological constant we have

$$
N(r)=\sqrt{\left( m+\frac{2}{3c_{1}}
\frac{r^{2}}{l^{2}}+\frac{1}{9c_{1}^{2}}%
\frac{k^{2}}{r^{2}}\right) }. \eqno(3.25)
$$

Let us now investigate the time evolution of $g_{21}.$
Substituting the quantity $P_{21} = r^{2}f^{-2}P^{21}$
in the Eq. (3.11)  we find

$$
C(r)=-3c_{1}\partial _{r}f(r)\mp
\frac{2}{l^{2}}f(r)^{-1}r-\frac{rf(r)^{-1} %
}{3c_{1}}\left[ P^{21}(r)\right] ^{2},
$$
from what we obtain

\begin{eqnarray}
\dot{g}_{21}(r,t) &=&\frac{\delta H}{\delta P^{21}(r)}\nonumber \\
&=&\frac{\delta }{\delta P^{21}(r)}
\int dr^{^{\prime }}\biggl[ N^{\theta
}(r^{^{\prime }})
\biggl(-r^{^{\prime }}\partial _{r^{^{\prime }}}\left[
r^{^{\prime }}P^{21}(r^{^{\prime }})\right]  \nonumber \\
&&-r^{^{\prime }}P^{21}(r^{^{\prime }})\biggr) +N(r^{^{\prime
}})\biggl(-3c_{1}\partial _{r^{\prime}} f(r^{\prime})\mp  \nonumber  \\
&&\mp \frac{2}{l^{2}}f(r^{\prime})^{-1}r^{\prime}
-\frac{r^{\prime}f(r^{\prime})^{-1} }{3c_{1}}\left[
P^{21}(r^{\prime})\right] ^{2}\biggr)\biggr]. \nonumber
\end{eqnarray}
Taking into account that the solution
$P^{21}(r)=\frac{k}{r^{2}}$ is  time independent
and that  $\dot{g}_{21}(r,t)=0$ we arrive at

$$
-\partial _{r}N^{\theta }(r)r^{3}-\frac{2}{3c_{1}}k=0,
$$
whose solution is given by

$$
N^{\theta }(r)=\frac{1}{3c_{1}r^{2}}k.\eqno(3.26)
$$
We obtain again a result that is identical with the corresponding
quantity of Ref. \cite{14} after making $c_{1}^{{}}=-\frac{2}{3}$
and identifying $k=J$.

The event horizon of the black hole is determined by a surface of
constant radius $r=r_{H}$. The latter follows from the condition

$$
g_{00}=0=N(r)=f(r),  
$$
which implies

$$
r^{4}-\frac{3}{2}c_{1}l^{2}mr^{2}-\frac{1}{6c_{1}}l^{2}k^{2}=0.
$$
The solution is given by

$$
r_{\pm }=l\sqrt{\frac{3}{4}c_{1}m}\left[ 1\pm
\left( 1+\frac{8}{27}\frac{1}{%
c_{1}^{3}}\left(
\frac{k}{ml}\right) ^{2}\right) ^{1/2}\right] ^{1/2}.
\eqno(3.27)$$
By making $c_{1}=-\frac{2}{3}$ we obtain

$$
r_{\pm }=l\sqrt{-\frac{m}{2}}\left[ 1\pm
\left( 1-\left( \frac{k}{ml}%
\right) ^{2}\right) ^{1/2}\right] ^{1/2}, \eqno(3.28)
$$
which coincides with the radii  of the event horizons of the
BTZ black hole. The event horizons $r_{\pm}$ require a further
condition, namely, that  $m<0,$ or $m=-M$, where $M>0$ is the
mass of the black hole. Real solutions are obtained by requiring

$$
\left| k\right|  \leq \left| ml\right|,\eqno(3.29)
$$

$$
\left| J\right|  \leq \left| Ml\right|.\eqno(3.30)
$$

Therefore we have

\begin{eqnarray*}
M &>&0\Longrightarrow {\rm black{\;} hole{\;} solution,} \\
M &=&0,\;\;J=0\Longrightarrow {\rm Absence{\;} of{\;} black
{\;}hole,} \\
M &=&-1,\;\;J=0\Longrightarrow {\rm Anti-de{\;} Sitter{\;}
solution.}
\end{eqnarray*}

\bigskip
\bigskip

\noindent{\bf \S 4. Gravitational energy in 2+1 dimensions}\par
\bigskip

A natural application of the present formalism consists in
calculating the gravitational energy of black holes in 2+1
dimensions. The Hamiltonian constraint Eq. (2.12) contains a
total divergence. In analogy  to the 3+1 case, the total
divergence given by $-\partial _{k}\left( 3c_{1}eT^{k}\right)$
(apart from a multiplicative constant) is interpreted
as the gravitational energy density, and the integral form of the
Hamiltonian constraint is likewise interpreted as an energy
equation for the gravitational field. Therefore the
gravitational energy is defined as a surface integral,

$$E=-{1\over{16\pi G}}\oint_C dS_k(3c_1eT^k)-
E_0\;,\eqno(4.1)$$

\noindent where $C$ is, in the present case, a one-dimensional
closed contour, and $E_0$ is a suitable reference energy value.

By requiring $C$, defined by the condition $r=r_0$, to be a
closed contour of a surface $S$, we find that in the present case
the energy contained within the interior of $S$ is given by 

$$E=-{1\over{16\pi G}}\int_C d\theta(3c_1eT^1)\,\,-E_0$$

$$={{3c_1}\over{ 2 \pi}}\int_C d\theta f(r)\,\,-E_0=
{{3c_1}\over{2 \pi}}\int_C d\theta\,
\sqrt{\left( m-\frac{2}{3c_{1}}
\frac{r_0^{2}}{l^{2}}+\frac{1}{9c_{1}^{2}
}\frac{k^{2}}{r_0^{2}}\right) }\,\,-E_0\;.\eqno(4.2)$$

\noindent where we have made $G=1/8$.
We remark that in previous analyses of the gravitational energy
of black holes\cite{6,7}, the latter expressions were evaluated
by means of a surface integral. This approach circumvents the
problem of dealing with imaginary field quantities. Moreover
we note that the coordinate system employed above is not well
defined inside the event horizon (a similar situation takes place
in the Schwarzschild and Kerr space-times, described by the usual
spherical coordinates\cite{6,7}).

We will consider now the metric corresponding to the rotational
black hole with $\Lambda <0$ and will compare it with
the result that has been obtained in the literature. For this
purpose we will fix the value of $c_1$ according to
$c_1=-2/3$. The energy contained within the circle of constant
radius $r_{0}$ reads

$$
E=-\frac{1}{\pi }\int_{0}^{2\pi }d\theta
f(r) -E_0 =-2\left( \sqrt{\left( -M+
\frac{r_{0}^{2}}{l^{2}}+\frac{1}{4}\frac{%
J^{2}}{r_{0}^{2}}\right) }-\frac{r_{0}}{l}\right),\eqno(4.3)
$$
where we have set $E_0= -2(r_{0}/l)$, $m=-M$ and $k=J$. We follow
the interpretation of Ba\~{n}ados {\it et al.}, according to
which the vacuum state is obtained by requiring the mass of the black
hole to vanish, which, by Eq. (3.30), implies
the vanishing of the angular momentum, $J=0$. Therefore for the
vacuum energy we have $E=0$.
Note that the black hole energy also vanishes if we extend
the integration to the whole bi-dimensional space.
The energy value $E=0$ corresponds to a space-time with the line
element given by

$$
ds_{vac}^{2}=-\left( \frac{r}{l}\right) ^{2}dt^{2}
+\left( \frac{r}{l}\right)^{-2}dr^{2}+r^{2}d\theta ^{2}.\eqno(4.4)
$$

Let us now show that we can obtain the result above out of 
different diads. We will consider the diads

$$
e_{\left( k\right) i}=\left(
\begin{tabular}{ll}
$\alpha \cos \theta $ & $-r\sin \theta $ \\ 
$\alpha \sin \theta $ & $r\cos \theta $%
\end{tabular}
\right),\eqno(4.5)
$$
where $f(r)^{-1}=\alpha =\left( -M+\frac{r^{2}}{l^{2}}
+\frac{1}{4}\frac{J^{2}%
}{r^{2}}\right) ^{-1/2}.$

The bi-dimensional metric tensor $g_{ij}$ has the nonvanishing
components $g_{11} =\alpha^{2},\;\; g_{22} =r^{2}$,
and $e=\det \left(e_{\left( k\right)i}\right)=r\alpha$.
The nonzero components of torsion tensor are

\begin{eqnarray*}
T_{\left( 1\right) 12} &=& (\alpha -1) \sin \theta, \\
T_{\left( 2\right) 12} &=& -(\alpha -1)\cos \theta.
\end{eqnarray*}
Note that these components vanish by making $\alpha =1$.
After some computations we obtain

$$
eT^{1}=\left( 1-\alpha ^{-1}\right),\;\;\;eT^{2}=0.
$$
Hence

$$
\partial _{i}\left( eT^{i}\right)
=\partial _{r}\left( 1-\alpha ^{-1}\right).\eqno(4.6)
$$

Requiring again  $c_{1}=-\frac{2}{3}$ and making $G=1/8$ we find

$$
E =-\frac{1}{16\pi G}\int_{S} dS_k
\left( 3c_{1}eT^{k}\right)-E_0
=\frac{1}{\pi }\int_{0}^{2\pi }d\theta \left( 1-\alpha
^{-1}\right)-E_0
$$

$$=2\left( 1-\alpha ^{-1}\right) -E_0
=-2\left[ \left( -M+\frac{r_{0}^{2}}{l^{2}}
+\frac{1}{4}\frac{J^{2}}{%
r_{0}^{2}}\right) ^{1/2}-\frac{r_{0}}{l}\right] , \eqno(4.7)
$$
where $E_0=2(1-r_0/l) $ was adjusted to obtain $E=0$ when
making $M=J=0$.

The value above for the gravitational energy of the BTZ black hole
is identical to the value obtained by means of the Brown-York
method, according to Refs.  \cite{16,17},

$$
E_{BY}=-2\left( \sqrt{-M+\frac{r_{0}^{2}}{l^{2}}
+\frac{1}{4}\frac{J^{2}}{%
r_{0}^{2}}}-\varepsilon_{0}\right) ,\eqno(4.8)
$$
where the reference energy $\varepsilon_{0}$ is, in principle,
arbitrarily choosen. In Refs. \cite{16,17} the reference energy is
taken to be the energy of the Anti-de Sitter space-time,

$$
\varepsilon_{0}=\sqrt{1+\frac{r_{0}^{2}}{l^{2}}},
$$
for which $M=-1$ and $J=0$. In this case, the vacuum
line element is taken to be

$$
ds^{2}_{vac}=-\left[ 1+\left( \frac{r}{l}\right) ^{2}
\right] dt^{2}+\left[
1+\left( \frac{r}{l}\right) ^{2}\right] ^{-1}
dr^{2}+r^{2}d\theta ^{2}.
$$

In Ref. \cite{16} it is also adopted  a reference energy that
corresponds to the absence of black holes, $M=J=0$, 

$$
\varepsilon_{0}=\frac{r_{0}}{l}.
$$
We may obtain the reference energies above in our
expression (4.7) since we  may conveniently
adjust the constant of integration.

We observe that we may establish an alternative possibility for
the reference energy by taking the reference space-time to be the
Minkowsky space-time. In this case we must have $M=-1,$ $J=0$ and
$ 1/l^{2}\rightarrow 0$, or taking $\alpha =1$,

$$E_{0}=-2.$$
After taking the values above for $M,\,J$ and $ 1/l^{2}$
the line element reduces to

$$
ds^{2}_{vac}=-dt^{2}+dr^{2}+r^{2}d\theta ^{2}.
$$

Therefore we conclude that the energy expressions
(4.3) and (4.7) obtained in the context of the 2+1 teleparallel
formulation yield consistent results which were previously
obtained in the literature.
\bigskip
\bigskip

\noindent{\bf \S 5. Conclusions}\par
\bigskip

In this work we investigated the existence of black holes in
2+1 teleparallel theories of gravity. The analysis has been
carried out in the realm of the  Hamiltonian formulation of
arbitrary, quadratic teleparallel theories with a negative
cosmological constant. We considered
gravitational fields with circularly
symmetric geometry and obtained a black hole solution
of the Hamiltonian field equations that generalizes the BTZ black
hole. The latter is obtained by fixing the free parameter
of the theory. We also investigated the gravitational energy
of this solution by means of the simple expression first
analyzed in the context of 3+1 teleparallel theories. The energy
expressions are obtained in a straightforward way and are
consistent with the results previously obtained by means
of the Brown-York method. We observed that there are various
possibilities for fixing the reference space-time. The
simplicity of the mathematical structure of the 2+1 dimensional
formulation motivates us to investigate the notion of gravitational
angular momentum in this framework.
This issue is currently being studied.\par

\bigskip
A. A. S. is grateful to CNPq (the Brazilian National
Research Council) for the financial support.\par

\end{document}